\documentclass[aps,prd,showkeys,showpacs,preprint,11pt]{revtex4-1}
\usepackage{amsmath,amsthm,amsfonts,amssymb}
\usepackage[mathcal]{eucal}
\usepackage{mathrsfs}
\usepackage{graphicx}
\usepackage{dcolumn}
\usepackage{latexsym}
\usepackage{epsfig}
\usepackage{bm}
\usepackage[all]{xy}
\usepackage[english]{babel}
\usepackage{amsmath,amsthm,amsfonts,amssymb}
\usepackage[mathcal]{eucal}
\usepackage{mathrsfs}
\usepackage{natbib}
\usepackage{lmodern}
\usepackage{graphicx}
\usepackage{dcolumn}
\usepackage{latexsym}
\usepackage{epsfig}
\usepackage{bm}
\usepackage[all]{xy}
\usepackage[utf8]{inputenc}
\usepackage{scalefnt}
\usepackage{dsfont}
\usepackage{xcolor}
\usepackage{hyperref}
\hypersetup{
	colorlinks,
    citecolor=blue!20!green!90!black!99,
    filecolor=red,
    linkcolor=blue!20!red!90!black!99,
    urlcolor=gray
    }

\DeclareMathOperator{\e}{e}
\DeclareMathOperator{\x}{x}

\begin{document}


\title{Configurational entropy and Newton's Law in double sine-Gordon braneworlds}

\author{W. T. Cruz $^{1}$}
\email{wilami@ifce.edu.br}
\author{D. M. Dantas $^{2}$}
\email{davi@fisica.ufc.br}
\author{R. V. Maluf $^{2}$}
\email{r.v.maluf@fisica.ufc.br}

\author{C. A. S. Almeida $^{2}$}
\email{carlos@fisica.ufc.br}

\affiliation{$^{1}$Instituto Federal de Educa\c{c}\~ao, Ci\^encia e Tecnologia do Cear\'a (IFCE),
Campus Juazeiro do Norte, 63040-000 Juazeiro do Norte - CE - Brazil}
\affiliation{$^{2}$ Universidade Federal do Cear\'a (UFC), Departamento de F\'isica, Campus do
Pici, Fortaleza - CE, C.P. 6030, 60455-760 - Brazil}

\begin{abstract}
  In this work, we evaluate the Shannon-like entropic measure of spatially-localized functions for a five-dimensional braneworld generated by a double sine-Gordon (DSG) potential.  The   {differential configurational
entropy (DCE)} has been shown in several recent works to be a   {configurational informational measure (CIM)} that selects critical points and brings out phase transitions in confined energy models with arbitrary parameters. We select the DSG scenario because it presents an energy-degenerate spatially localized profile where the solutions to the scalar field demonstrate critical behavior that is only a result of geometrical effects.  As we will show, the DCE evaluation provides a method for predicting the existence of a transition between the phases of the domain wall solutions. Moreover, the entropic measure reveals information about the model that is capable of describing the phase sector where we obtain resonance modes on the massive spectra of the graviton. The graviton resonance lifetimes are related to the existence of scales on which 4D gravity is recovered. Thus,  we correlate the critical points defined by the   {CIMs} with the existence of resonances and their lifetimes. To extend our research regarding this system, we calculate the corrections to Newton's Law coming from the graviton modes.  
\end{abstract}

\maketitle

\section{Introduction}
\label{intro}

The term Configuration Entropy (CE) was reintroduced by Gleiser and Stamatopoulos in 2012 \cite{gleiser1}, based on Shannon's information theory \cite{shannon}. In this view, the CE represents a measure of stability applied to a spatially bound energy system. To demonstrate the applicability of CE, early works \cite{gleiser1,gleiser-plb,gleiser-prd} show how this measure can be used to find stable Q-balls and the Chandrasekhar limit for white dwarfs \cite{gleiser-plb}. Furthermore, the CE was applied to investigate the nonequilibrium dynamics of spontaneous symmetry-breaking \cite{gleiser-prd} and also to obtain bounds on the stability of various self-gravitating astrophysical objects  \cite{gleiser-prd2}. Subsequently, several papers have been released where the CE was applied in various areas of physics\cite{ce1,ce2,ce3,ce4,ce5}, as topics in  AdS/QCD holography \cite{ce1,ce2}, the Korteweg–de Vries equation \cite{ce3}, Bose-Einstein condensates \cite{ce4}, and predicting atomic decay rates \cite{ce5}.   {More recently, configurational information measures (CIMs) were extended to include the concepts of differential configurational entropy (DCE) and differential configurational complexity (DCC) \cite{n1,n2}. In our approach, to study the dynamics of phase transitions, we consider the DCE definition.}

In the present paper, we focus on recent application of   {CIMs} to braneworld models \cite{ce-b1,ce-bloch,ce-weyl,ce-6d,ce-sg,ce-mariana}. In fact, the CE has been shown to be useful for determining stable parameters on brane models with a single scalar field configuration, as in the case of  $F(R$) brane models \cite{ce-b1}, the Bloch brane \cite{ce-bloch}, Weyl's brane \cite{ce-weyl}, string-like models \cite{ce-6d}, braneworld scenarios from deformed defect chains \cite{ce-mariana}, and the (single) sine-Gordon model \cite{ce-sg}. As another interesting feature, the CE also determines where phase transitions occur in the double-kinks models \cite{ce-bloch,ce-weyl}, as in the case of some degenerate branes \cite{ce-bloch}. The kink to multi-kink phase transitions have some important applications. In general, multi-kink models tend to preserve stability, whereas single-kink models are no longer stable \cite{multi1}, in addition, the multi-kink models have some specific experimental applications \cite{multi2,multi3}. Moreover, in the braneworld context, multi-kink models allow the appearance of multiple resonant states \cite{dsg-gravi,dsg-gauge}.

 In this paper, we aim to explore the configuration of the double sine-Gordon (DSG) model.  The DSG model represents a version of the sine-Gordon model where the scalar field is a double-kink, instead of one single kink. We are especially interested in the DSG setup because it presents a smooth transition between the usual kink solution and the two-kink structure. In this work, we propose a new kind of DSG potential that allows the convergence from the two-kinks DSG model to a single-kink model, similar to the sine-Gordon (SG) scenario, through the parameter $a$. From previous works \cite{ce-bloch}, we observe that the phase transitions in brane solutions dynamically generated by scalar fields are predicted by the existence of a minimum on the Shannon-like entropic measure of the energy density. Because the DSG model possesses a phase transition in the scalar field solution, we expect to obtain the same behavior for the   {DCE} measure, that is, a minimum on the   {DCE}  related to the phase transition.

After mapping the DSG setup regarding the   {DCE} measure, we proceed to investigate graviton localization. In our previous work \cite{dsg-gravi}, we have shown the existence of resonances on the massive spectrum of the graviton in the DSG setup. The existence of such resonances is related to the existence of a long-distance scale where the Newtonian potential is valid \cite{csaki2,metastable}. Moreover, we calculate the corrections to Newton's potential and its correlations with the resonant modes. Additionally, the localization of fields such as gravity \cite{dsg-gravi} and the vector gauge-fields and the correction to Coulomb's Laws \cite{dsg-gauge} have been performed on a similar DSG model.

This work is organized as follows. In Sec. \ref{sec:brane}, we review the DSG brane scenario, where we show that the DSG model can can lead to the simple sine-Gordon model by a suitable choice of parameters. In Sec. \ref{sec.entropy}, our first results are shown. We apply the   {DCE} method to obtain the critical points for the $a$ parameter of DSG model. In Sec. \ref{sec-results}, we detail the gravity localization on the DSG model in subsection \ref{sec:review-gravi}, and the resonant modes in subsection \ref{sec:res}. The corrections to Newton's law using the DSG model are computed in subsection \ref{sec:newton}. Finally, we summarize our results and give final comments in Sec. \ref{sec.conc}.

\section{Double sine-Gordon branes}
\label{sec:brane}

To construct our braneworld, we first consider a scalar field coupled to gravity defined by the following action
\begin{equation}\label{eq:action}
S=\int d^{5}x\sqrt{-G}\Big[\frac{1}{4}R-\frac{1}{2}\partial_{\mu}\phi\partial^{\mu }\phi-V(\phi)\Bigr].
\end{equation}
The scalar field depends only on the extra dimensional coordinate $y$ on which space-time is asymptotic $AdS_{5}$ with the metric $ds^{2}=e^{2A(y)}\eta_{\mu\nu}dx^{\mu}dx^{\nu}+dy^{2}$, where $A(y)$ is the warp factor and $R$ is the scalar curvature   {in five dimensions}.  For the  metric $\eta_{\mu\nu}$ with $\mu,\nu=0,1..3$, we use the signature $(-,+,+,+)$. 

The resulting equations of motion are \begin{equation}
\phi'^{2}-2V(\phi)  =  6A'^{2}\label{eq:eqmov1}
\end{equation}
\begin{equation}
\phi'^{2}+2V(\phi)  =  -6A'^{2}-3A''\label{eq:eqmov2}
\end{equation}
\begin{equation}
\phi''+4A^{\prime}\phi'  =  \frac{\partial V}{\partial \phi}.
\label{eq:eqmov3}
\end{equation}

  {At this point, it is convenient to adopt a very useful first-order formalism to obtain solutions to the equations of motion. This technique appeared first in the study of supergravity domain walls and has been widely used for various thick brane models  \cite{bazeia1,bazeia3,shif,alonso,de,cvetic}.  Writing the ordinary potential $V(\phi)$ in terms of a $W(\phi)$ superpotential function 
\begin{equation}
\label{eq:vsup} V(\phi)=\frac{1}{8}\left(\frac{\partial W}{\partial\phi}\right)^{2}-\frac{1}{3}W^{2},
\end{equation}
and setting $\phi^{\prime}=\frac{1}{2}\frac{\partial W}{\partial\phi}, \ \ \ A^{\prime}=-\frac{1}{3}W$, yields a solution. A specific choice of the $W(\phi)$ function results in the required scalar field potential (SG or DSG) and also defines the solutions to $\phi(y)$ and $A(y)$.}

The SG potential is achieved with the superpotential function by $W(\phi)=3 b c \sin\left(\sqrt{\frac{2}{3 b}\phi}\right)$ \cite{dsg-gravi}. However, we are now interested in a more general class of potentials that can provide us two-kink solutions \cite{dsg3,dsg4,aplications,dsg_ap2}. Such a potential is named double sine-Gordon (DSG) because it has a relative minimum at $\Phi=\pm\pi$ beyond the absolute minimum at $(0,2\pi)$, resulting in double bounce solutions to the scalar field. The change in the original vacua of the SG model gives us a new scalar field solution, interpolated between the two vacua ($ 0$ and $2\pi$) with a transient state at $\pi$ \cite{dsg1,dsg2,dsg3,dsg4}.

Starting from action (\ref{eq:action}), we propose a new super potential to find a   {new solution capable of interpolating both DSG and SG models}
\begin{eqnarray}\label{eq:wdsg2}
{W}(\Phi)= 4 \cos(\Phi/2) \sqrt{a + 2\big(1+ \cos(\Phi)\big)} \nonumber\\ + 2 a \ln\left[2 \cos(\Phi/2) + \sqrt{a + 2\big(1+ \cos(\Phi)\big)}\right],
\end{eqnarray}
which result in the new modified DGS potential shown in Eq. (\ref{eq:v_dsg}).
\begin{figure*}\begin{eqnarray}\label{eq:v_dsg}
V(\Phi)=\frac{1}{8}\left(\frac{\partial \mathcal{W}}{\partial\Phi}\right)^{2}-\frac{1}{3}\mathcal{W}^{2} = 1 -\cos(2\Phi) + a\big(1 - \cos(\Phi)\big) - \frac{5}{4}\Bigg[2 \cos(\Phi/2) \sqrt{a + 2\big(1+ \cos(\Phi)\big)} \nonumber\\
+a \ln\left(2 \cos(\Phi/2) + \sqrt{a + 2\big(1+ \cos(\Phi)\big)}\right)\Bigg]^2.
\end{eqnarray}
\end{figure*}

\begin{figure*}
\centering
\resizebox{0.8\textwidth}{!}{
\includegraphics{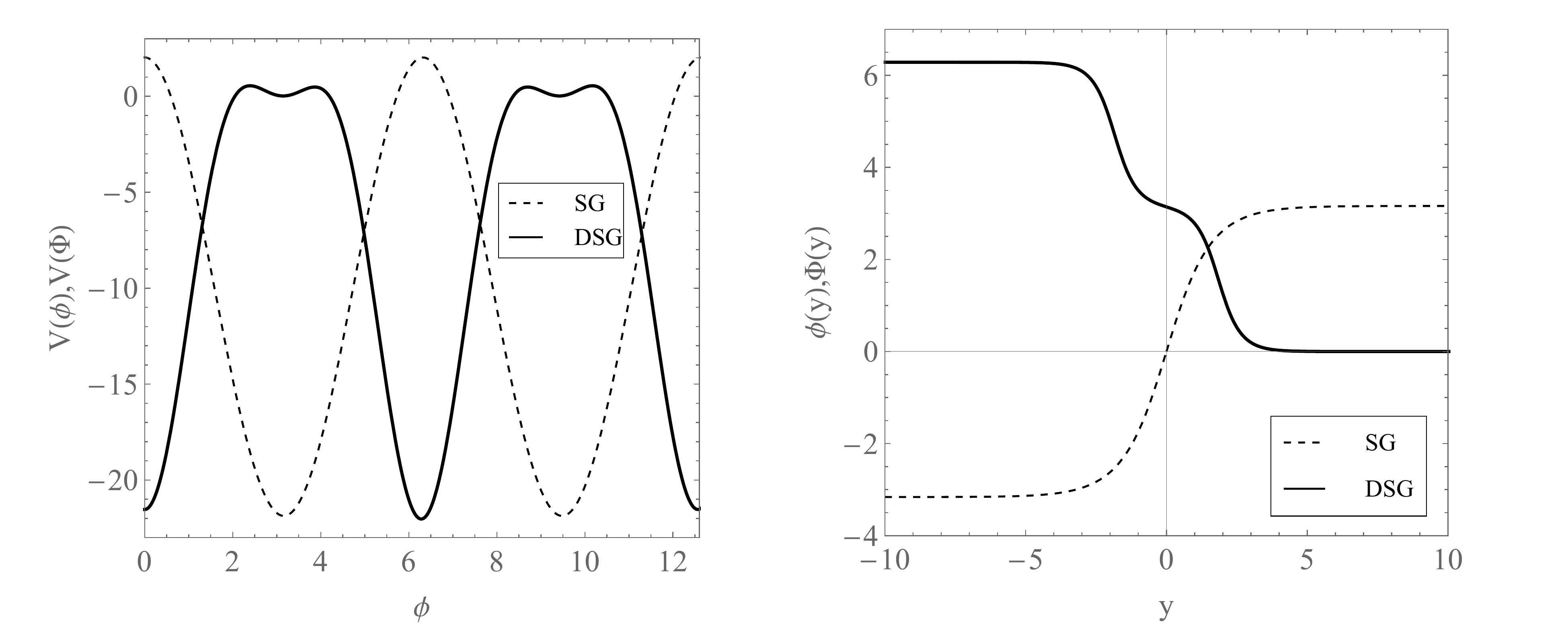}}
 \caption{Plots of SG and DSG potentials (left). Plots of scalar field solutions (right). }
\label{fig:sg1}      
\end{figure*}

We denote by $\Phi(y)$ the scalar field   {in the DSG setup} obtained  numerically from the first-order equation  $\Phi^{\prime}=\frac{1}{2}\frac{\partial {W}}{\partial\Phi}$. In Figure  \ref{fig:sg1}, we plot the SG and the DSG potentials with their respective solutions to the scalar field. The minima of the SG potential corresponds to the vacua of $\phi(y)$; however, in the DSG scenario, there is a local minimum at $\Phi=\pi$ beyond the minima at $\Phi=2 \pi$. The existence of the local minima results in the structure of  a double bounce. 

In fact, it is possible to obtain a single kink solution from the DSG potential. The structure of the scalar field solution with two bounces transitions to a kink-like solution. This transformation coincides with the vanishing of the transient minimum in the DSG potential at $\Phi=\pi$.

\begin{figure*}[h]
\centering
\includegraphics[width=0.8\textwidth]{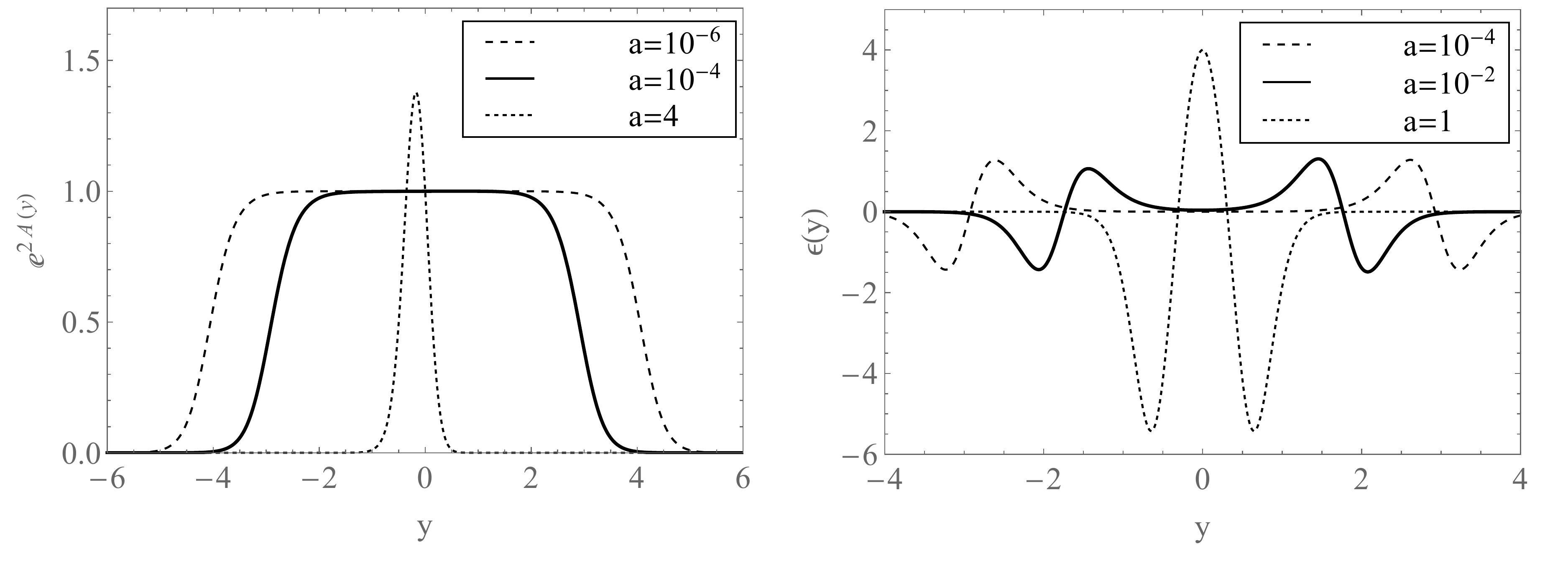}
\caption{Warpfactors $\e^{2A(y)}$ (left panel) and energy density $\varepsilon(y)$ (right panel) in the DSG model.}\label{fig:energy}
\end{figure*}

The   {new DSG} solution for the warp factor is obtained from equation 
\begin{equation}
{A}'(y)=-\frac{1}{3}{W}(\Phi).  \label{warp-factot-dsg}
\end{equation}
We plot the numerical solutions of Eq. \eqref{warp-factot-dsg} in the left panel of Figure  \ref{fig:energy}, where we note that an increase of $a$ narrows the warp-factors.

Finally, the energy density   {for the DSG} is given by:
\begin{equation}
\varepsilon (y)=\e^{2{A}(y)}\left[ \frac{1}{2}\Phi ^{\prime 2}+V(\Phi)\right] .  \label{en0}
\end{equation}
This expression is shown in the right panel of Figure  \ref{fig:energy}, where we note that the increase in $a$ confines the energy. The computation of the configurational entropy is based on this energy density.

Here, we finish the review of the DSG model. In the next section, we will obtain our first result when applying the configurational entropy to the energy density of the DSG model. 

\section{Configurational entropy}\label{sec.entropy}

\begin{figure*}
\centering
\includegraphics[width=0.8\textwidth]{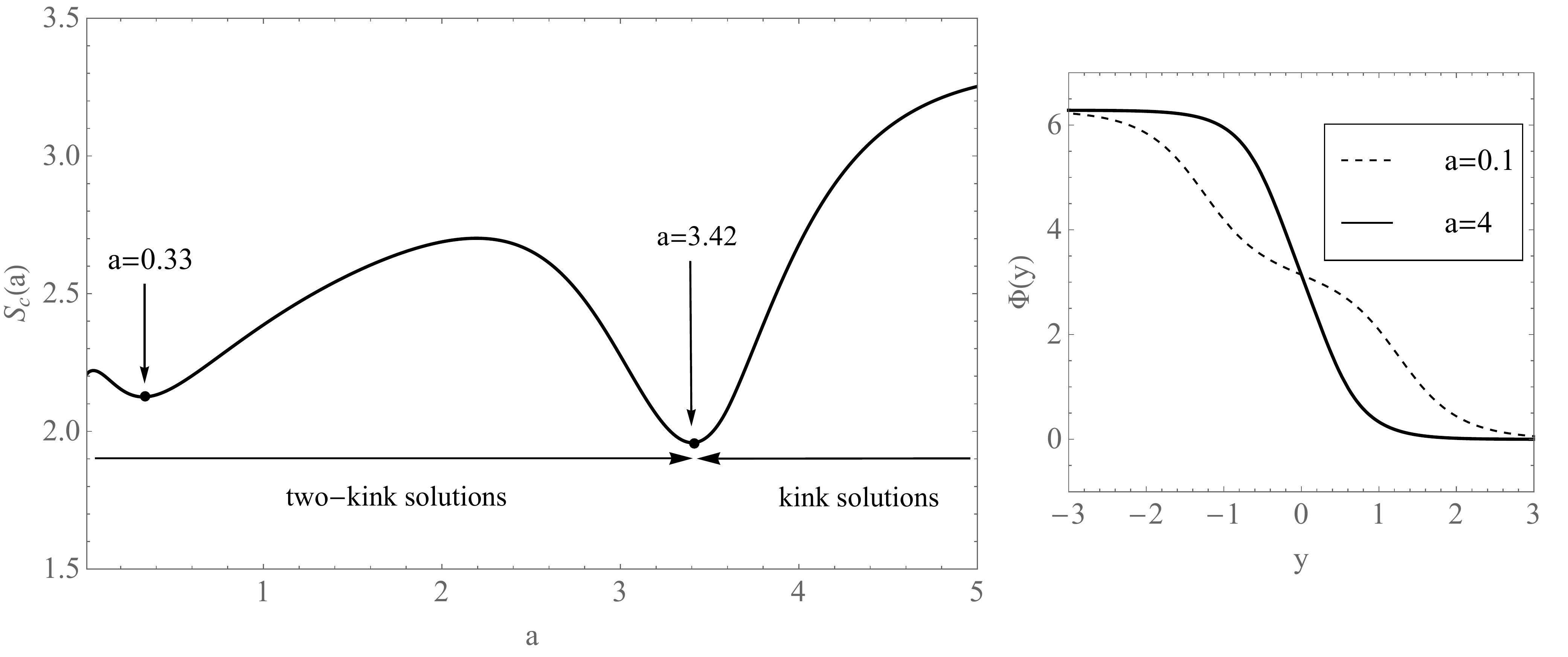}
\caption{{ Plots of the   {DCE} (left) and the scalar field solution for the DSG setup (right).} }\label{fig:sg2}
\end{figure*}

  {The DCE is related to the measure of the informational complexity of a localized field configuration and can be expressed by the Fourier transform of the energy density \cite{ce-b1,gleiser1}. This quantity was previously considered to study single sine-Gordon braneworlds \cite{ce-sg}, but the DSG structure was not computed.} 

  {The DCE is computed in terms of the probability density \cite{gleiser1,PLBgleiser-sowinski,PRDgleiser-stamatopoulos, n2} 
\begin{equation}
f(\omega )=\frac{{\lvert \mathcal{F}(\omega )\lvert ^{2}}}{{\int_{-\infty
}^{\infty }{d\omega' \lvert \mathcal{F}(\omega ')\lvert ^{2}}}\,},
\end{equation}
where $\mathcal{F}(\omega )=-\frac{1}{\sqrt{2\pi }}\int_{-\infty}^{\infty }{\varepsilon (y)\e%
^{i\omega y}dy}$ is the Fourier transform of the energy density Eq. \eqref{en0}.}

  {
Therefore, localized and continuous function ${f}(\omega )$ yields the
following definition for the DCE: 
\begin{equation}
S(f)=-\int_{-\infty}^{\infty }{d\omega }f(\omega )\mbox{ln}%
\left[ f(\omega )\right] .  \label{CE}
\end{equation}}

  {The DSG scenario is capable of creating multi-kink solutions. We expect to achieve a similar behavior to the   {CIMs} found in the study of degenerate Bloch branes \cite{ce-bloch}.}

  {We evaluate the DCE numerically, and the result is shown in Figure  \ref{fig:sg2}. The DCE evaluation reveals only positive values, excluding unphysical distributions. We observe two minima, at $a=0.33$ and $a=3.42$, with respective information-entropic measures $S_c=2.12$ and $S_c=3.42$. The presence of a local minima on the two-kink sector ($a=0.33$)  is due to the  DSG potential behaviour. For small values of $a$, there is a transient wave trough at $\Phi=\pi$ on the ${V}(\Phi)$ potential, as we have shown in Fig. \ref{fig:sg1}. This structure disappears as $a$ increases, giving rise to a new intermediate trough at $\Phi=2\pi$. In terms of the DCE measure, this effect is identified with the local minina at the two-kink sector.}

The DSG setup admits the usual kink and also the two-kink solution, as shown in Figure  \ref{fig:sg2}. By varying the parameter $a$, the scalar field transitions smoothly between these two structures, with a phase transition localized at the second stage.  The changing point from the kink to the two-kink solution occurs at the value of the parameter $a$  corresponding to the minima   {DCE} of the DSG braneworld. A similar result is found in the degenerate Bloch brane models\cite{ce-bloch}.

The parameter $a$ determines the shape of our brane solutions, as well as the warp factor and the energy density. The evaluation of the configurational entropy in terms of this variable allows us to predict the existence of a phase transition to the kink solution that is identified with minimum   {DCE}.  The results also show us that the minimum  {DCE} occurs in a sector where the energy density is more confined on the brane.  In the following sections, we proceed with  our investigation to connect the results above to the behavior of the graviton mass spectra and Newton's law.

\section{Gravity resonances and corrections to Newton's Law in the DSG model}\label{sec-results}

\subsection{Review of gravity localization in the DSG model}\label{sec:review-gravi}
We start by describing the conformally flat perturbed metric for our fifth dimension model: 
\begin{equation}
ds^2=e^{2A(y)}(\eta_{\mu\nu}+\epsilon h_{\mu\nu})dx^\mu
dx^\nu+dy^2, \label{eq:metric}
\end{equation}
where $A(y)$ is the warp factor Eq. \eqref{warp-factot-dsg}. $\eta_{\mu\nu}$ is the four-dimensional Minkowski metric. $h_{\mu\nu}=h_{\mu\nu}(x,y)$ represents the transverse and traceless ($\partial_{\mu}h_{\mu\nu}$ and $\eta^{\mu\nu}h_{\mu\nu}=\eta^{\mu}_{\nu}=0$) graviton, which has only 5 degrees of freedom, as appropriate for a spin-two 5D particle. This gauge imposition is common in 5D \cite{Csaki2, RS, Liu11}, but additional KK scalar/vectorial modes can be found in some 6D models \cite{Liu11}. $\overline{h}_{\mu\nu}$ can be described by the motion equation \cite{gremm,de}:
\begin{equation}\label{eq:grav}
\overline{h}_{\mu\nu}^{\prime\prime}+4A^{\prime}
\overline{h}_{\mu\nu}^{\prime}=e^{-2A}\partial^2\overline{h}_{\mu\nu}^{\prime},
\end{equation}
where $\partial^2$  is the four-dimensional wave operator and the primes denote derivatives with respect to the extra coordinate $y$.

To obtain a Schr\"{o}digner like equation for Eq. \eqref{eq:grav}, we need to transform the extra coordinate into the form $dz=e^{-A(y)}dy$. The metric of equation \eqref{eq:metric} becomes the conformal plane  $ds^2=e^{2A(z)}\left[(\eta_{\mu\nu}+\epsilon h_{\mu\nu})dx^\mu
dx^\nu + dz^2\right]$. Furthermore, performing the Kaluza-Klein coordinates decomposition $\overline{h}_{\mu\nu}(x,z)=e^{ip\cdot
x}e^{-\frac{3}{2}A(z)}U_{\mu\nu}(z)$, we can rewrite Eq. \eqref{eq:grav} as a
Schr\"{o}dinger-like equation, that describes the propagation of the gravitational modes into  the extra dimension $z$.
\begin{equation}
-\frac{d^2U(z)}{dz^2}+\mathcal{V}(z)\,U(z)=m^2\,U(z), \label{eq:schro}
\end{equation}
where the quantum-analogue potential reads
\begin{equation}
\mathcal{V}(z)=\frac{3}{2}\ddot{A}(z)+\frac{9}{4}\dot{A}^2(z),
 \end{equation}
where the dots denote derivatives with respect to $z$.

\begin{figure*}
 \centering
    \includegraphics[width=0.80\textwidth]{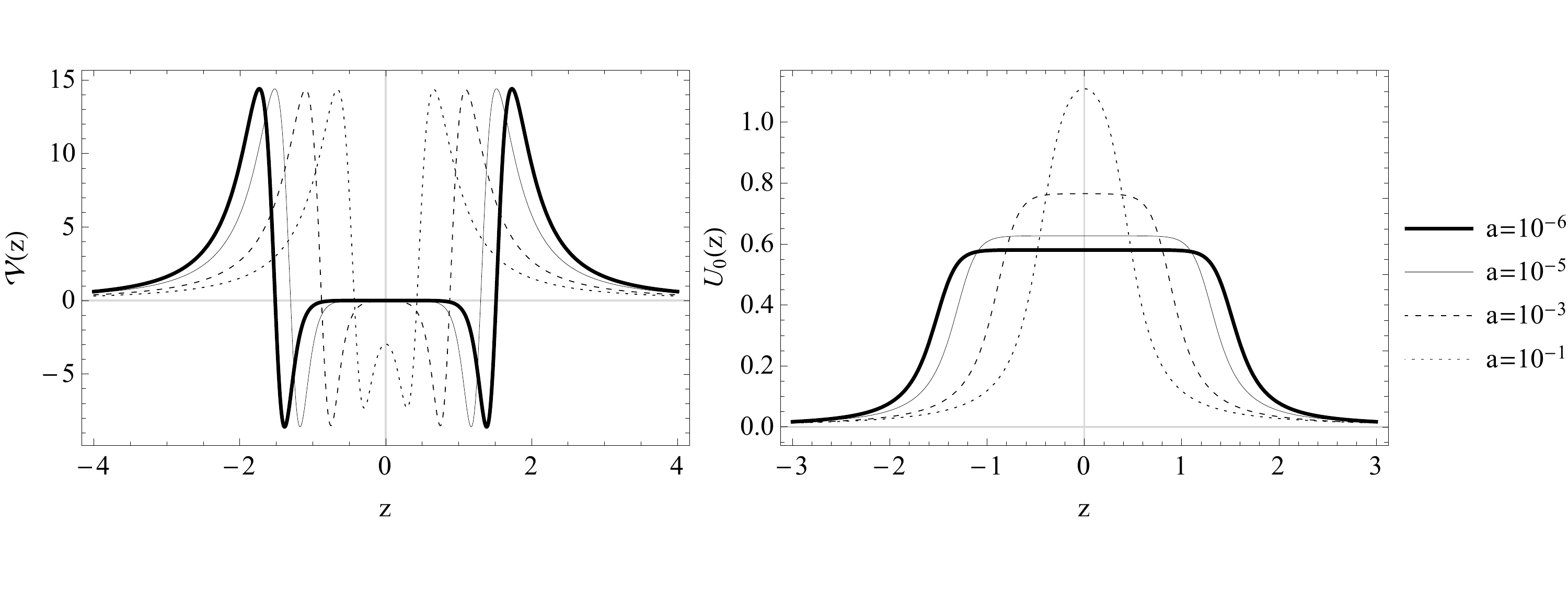}
 \caption{The Schr\"{o}edinger volcano-like potential for the DSG model (left panel) and the even normalized zero mode of the DSG model (right panel). The odd zero mode solutions have not been  localized in the DSG model.}
 \label{fig:quantumpotential}
\end{figure*}

 We plot, in the left panel of Figure  \ref{fig:quantumpotential}, the potential mentioned above, which assumes a volcano-like profile. This potential has two minima and two maxima, separated by the flat region around $z=0$. Note that the increase in the parameter $a$ narrows the width of the well; without reducing the amplitude of the maximum; this result suggests the presence of resonant states \cite{dsg-gravi}. For higher values of $a$, the usual volcano-like potential of the SG model is reproduced \cite{dsg-gravi}. It is also important to mention that the potential  $\mathcal{V}(z)$ supports no tachyonic states \cite{dsg-gravi}.

Moreover, as expected, equation \eqref{eq:schro} has unique localized (normalizable) massless mode solution \cite{dsg-gravi}. This zero mode is responsible for reproducing the four-dimensional gravity. We plot this zero mode in the right panel of Figure  \ref{fig:quantumpotential}. From here, we note, once more, that the increase in the parameter $a$ narrows the width of the zero modes. It is worth highlighting that both even and odd conditions are considered:
\begin{eqnarray}
\begin{tabular}{l l}
Even condition: $U(z=0)=cte$ and  $\dot{U}(z=0)=0$\\
Odd condition: $U(z=0)=0$ and  $\dot{U}(z=0)=cte$
\end{tabular}
\end{eqnarray}
However, for the DSG model,  it is only possible for the even solutions to be confined. Initially, $cte=1$, and its value is modified after wave normalization.

Conversely, the massive solutions of Eq. \eqref{eq:schro} are never normalizable, no matter if they are  an even solution or an odd solution. However, the modes can exhibit resonant states and also can perform corrections to the four-dimensional Newtonian potential. These aspects will be discussed in the next subsection. 

\subsection{Resonances in the DSG model}\label{sec:res}

 With the 3-brane placed at origin $z=0$, we can compute the coupling of the massive modes with the matter on the brane by solving the Schr\"{o}dinger-like equation \eqref{eq:schro}. Once the volcano-like potential $\mathcal{V}(z)$ represents only a small perturbation, we have the condition that the resonant mode should appear inside the well $m^2\leq \mathcal{V}_{max}$. Conversely, the massive solutions already acquire a plane wave profile after the barrier, when $m^2>\mathcal{V}_{max}$.

The so-called resonant modes are wavefunctions for which $U(0)$ has large amplitudes inside the brane (in comparison with its values far away from the brane). To find these resonant modes, we adopt the well-known relative probability method $N(m)$ \cite{chineses1,chineses2,chineses3,chineses4,chineses5} given by
\begin{equation}\label{rel_prob}
N(m)=\frac{\int_{-z_{b}}^{+z_{b}}|U_{m}(z)|^2 dz}{\int_{-z_{max}}^{+z_{max}}|U_{m}(z)|^2dz}.
\end{equation}
where $z_{max}$ denote the maximum box interval for $z$, while the narrow interval is given by  $z_b=0.1z_{max}$ \cite{chineses3}.

The function $N(m)$ is useful in the search for resonant modes. The term $\int_{-z_{b}}^{+z_{b}}|U_{m}(z)|^2 dz$ in Eq. \eqref{rel_prob} expresses the probability of finding the resonant modes in a narrow interval $-z_b<z<z_b$. The choice of the integration range $z_b=0.1z_{max}$ does not interfere with the masses of possible existing resonances \cite{nosso8}.

\begin{figure*}
 \centering
    \includegraphics[width=0.6\textwidth]{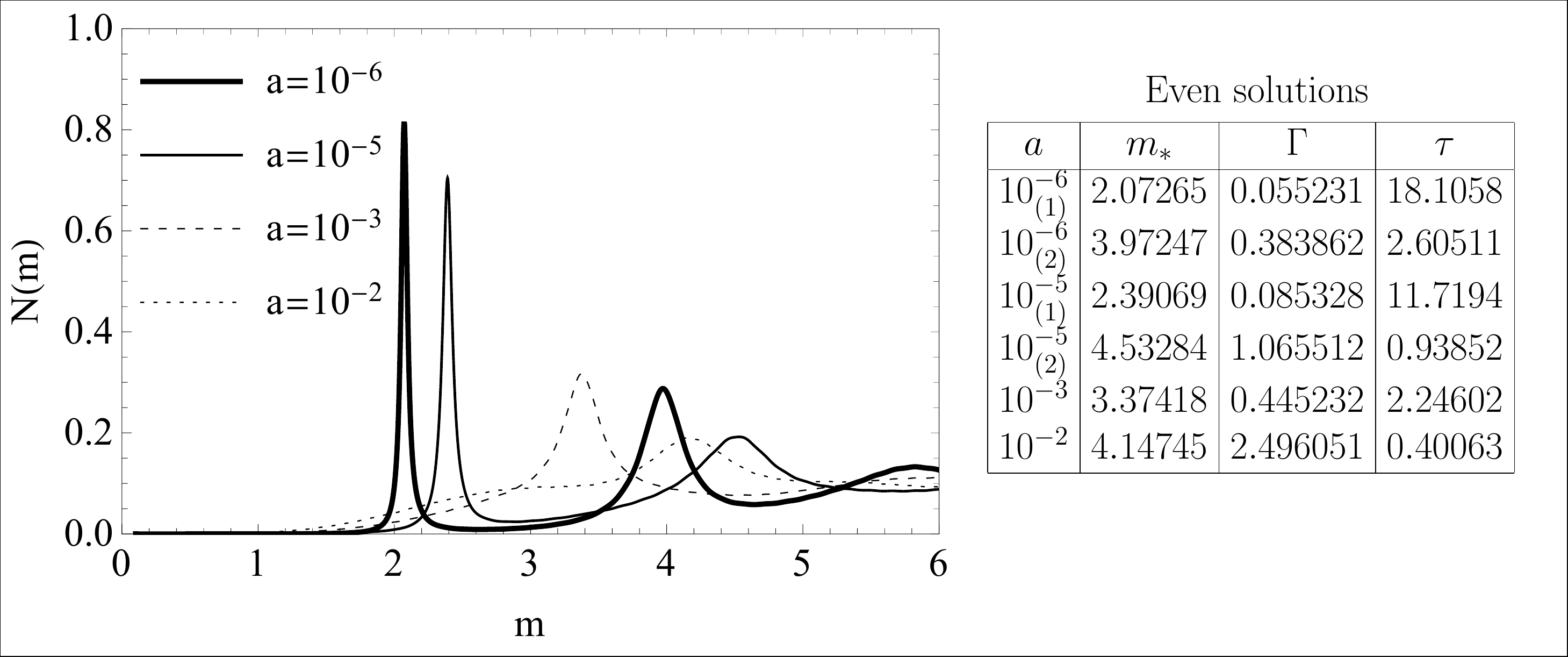}
 \caption{$N(m)$ method for gravity even solutions in the DSG model. $m_*$ is the resonant mass, $\Gamma$ is the the full width at half maximum  and  $\tau$ the lifetime. The labels $(1)$ and $(2)$ denote double peaks.}
 \label{fig:ress1a}
\end{figure*}

\begin{figure*}
 \centering
    \includegraphics[width=0.6\textwidth]{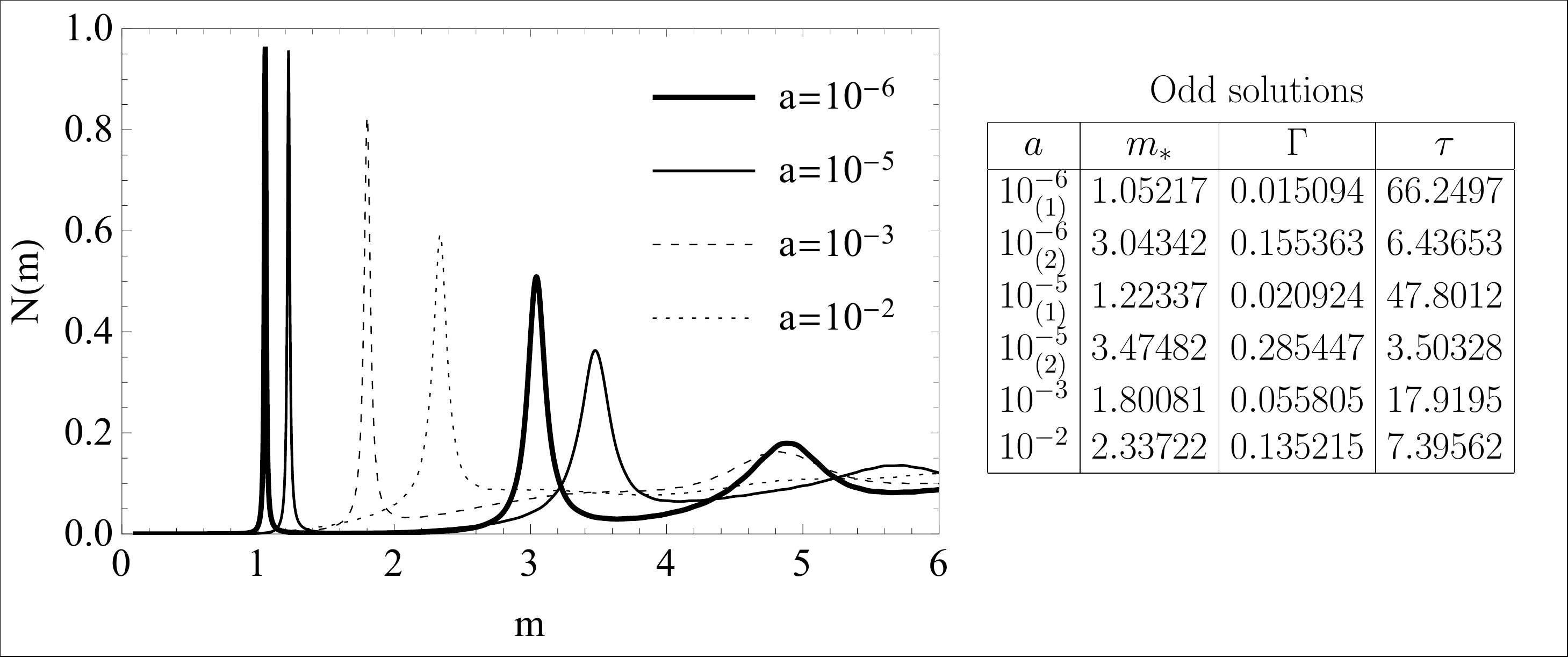}
 \caption{$N(m)$ method for gravity odd solutions in the DSG model. $m_*$ is the resonant mass, $\Gamma$ is the the full width at half maximum  and  $\tau$ the lifetime. The labels $(1)$ and $(2)$ denote double peaks.}
 \label{fig:ress2a}
\end{figure*}

We apply the relative probability method $N(m)$ from Eq. \eqref{rel_prob} to the DSG model for some values of parameter $a$. Figure  \ref{fig:ress1a} and Figure  \ref{fig:ress2a} were obtained for  even and odd gravity solutions, respectively. Each figure has an embedded table with the value of the  resonant mass $m_*$, the full width at half maximum $\Gamma$, and $\tau \approx \Gamma^{-1}$ is the lifetime. The labels $(1)$ and $(2)$ denote double peaks. 
 
From both Figure  \ref{fig:ress1a} and  Figure  \ref{fig:ress2a} it is clear that for $a=10^{-6}$ and $a=10^{-5}$ double peaks occur. By considering only the first peaks, we note that smaller $a$ leads to higher peaks, narrower $\Gamma$ and longer lifetimes $\tau$. Longer lifetimes implies more stable modes. In different circumstances, for larger $a$ the peaks vanish. Roughly, when $a>10^{-1}$, there are no more resonant modes because the peaks acquire the plane wave profile. This result corroborates the absence of resonant modes in single sine-Gordon models. Additionally, the even solutions have peaks placed at larger mass positions and have shorter lifetimes, compared with the odd solutions.

\begin{figure*}
 \centering
    \includegraphics[width=0.80\textwidth]{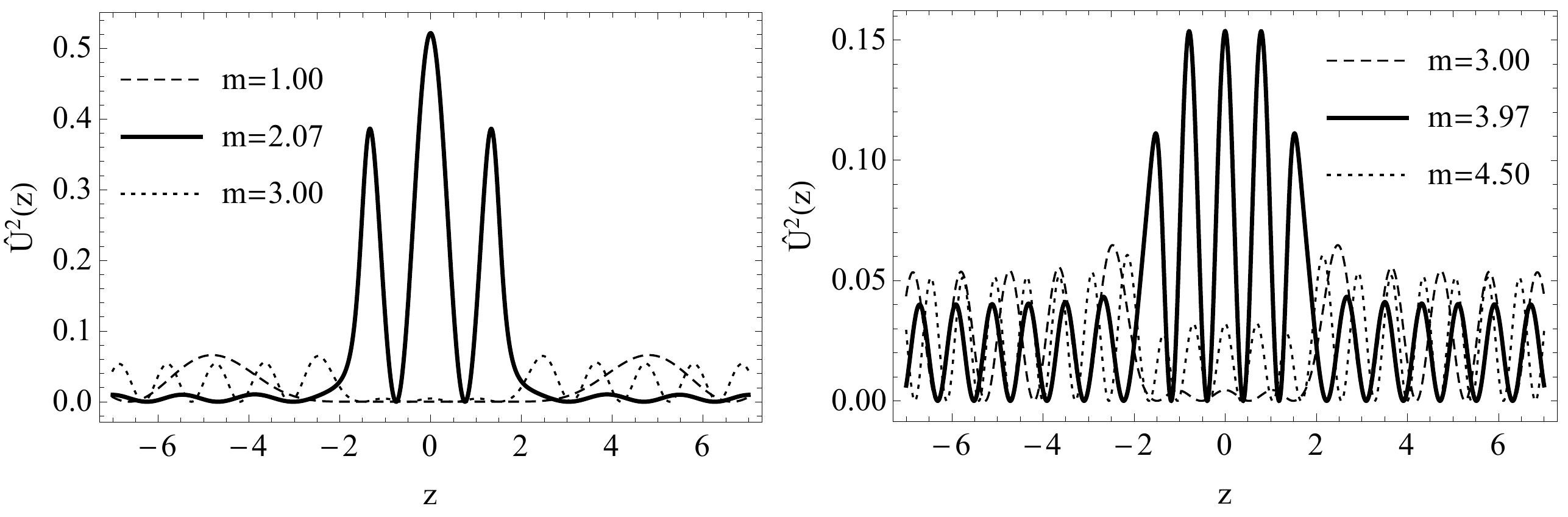}
 \caption{The profile for eigenfunction $\hat{U}^2(z)$ for some values of mass $m$. We use the even solution and $a=10^{-6}$. However, the feature of higher peaks for the resonant modes is present for all other solutions.}
 \label{fig:um}
\end{figure*}

 {Moreover, to clarify the gravity resonances, we plot the normalised profiles of $\hat{U}^2(z)$ for some values of mass $m$ in Figure \ref{fig:um}. Note that the resonant masses have much higher values than the non-resonant modes close to the origin (brane). This fact implies that the massive resonant modes are strongly coupled to the brane.}

 {Now, after proceeding with the analysis of the massive spectra we conclude that, in terms of the DCE, the existence of resonant modes is limited to the very beginning of the two kink sector.} The presence of resonances for odd or even solutions is limited to a very narrow interval $(a<10^{-1})$ of the parameter that controls the brane thickness.   Observing the structure of the volcano potential, we note that this is a region where the maxima of the potential are as far apart as possible. The entropic measure, corresponds only to the region at the beginning of the first stage interval, inside the two-kink solutions. However,  at the point where the phase transitions occur (DCE minimum) the solutions to the Schr\"{o}dinger-like equation acquire a plane wave structure, and there are no resonant modes. 

 {To improve our investigation, we must now define a CIM applied to quantify the complexity contained in the shape of the physical structures, known as the Differential Configurational Complexity (DCC) \cite{n1,n2}. A normalized modal fraction $\tilde{f}(\omega )$ specifies the ratio of the normalized Fourier transformed function and its maximum value ${f(\omega )}/{f_{max}}$.}

 {Therefore, localized and continuous function $\tilde{f}(\omega )$ yields the
following definition for the DCC:
\begin{equation}
C(\tilde{f})=-\int_{-\infty}^{\infty }{d\omega }\tilde{f}(\omega )\mbox{ln}%
\left[ \tilde{f}(\omega )\right] .  \label{CEb}
\end{equation}}
 {Large probability density amplitudes at the brane location obtained from the   Schr\"{o}dinger-like equation (\ref{eq:schro}) characterizes the resonant modes. A plane wave configuration must give vanishing DCE and DCC, as discussed in previous work \cite{n1}. However, analyzing the graviton massive spectra coming from Eq. (\ref{rel_prob}), for particular masses, modes with large $|U(z)|^2$ amplitude at $z=0$. So, the CIMs constructed from the probability density function should exhibit a particular behavior for the resonant modes are found. Conversely, modes with $m^2>\mathcal{V}_{max}$ tend to acquire a plane wave structure and must reach vanishing DCC and DCE. The evaluation of the CIMs with respect to the resonant modes are shown in  Table \ref{table} presenting an interesting relationship between their lifetimes and the corresponding DCC and DCE measures. In comparison with the plane wave solutions  $m^2>\mathcal{V}_{max}$, the resonant modes obtained for even and odd solutions exhibit higher values of DCC and DCE. We can observe such behavior from the values presented in Table \ref{table}, for $m=6$, the evaluated solutions present lower CIMs in comparison with the masses corresponding to the resonant modes. The results also indicate that the resonance lifetimes follow the CIMs magnitudes. Resonances with higher values toof DCE and DCC also lead to larger lifetimes.}

\begin{table}[]\centering\tiny
\begin{tabular}{|c|c|c|c|c|c|c|}
\hline
\multicolumn{4}{|c|}{Even solutions} & \multicolumn{3}{c|}{Odd solutions} \\ \hline
$m$    & $2.07265$  & $3.97247$  & $6.00000$       & $1.05217$   & $3.04342$   & $6.00000$   \\ \hline
DCE  & $1.84818$  & $1.07627$  & $0.30413$ & $1.86362$   & $1.6647$    & $0.40206$    \\ \hline
DCC  & $2.28046$  & $0.83229$  & $0.46257$ & $2.36965$   & $1.57937$   & $0.494147$   \\ \hline
$\tau$   & $18.1058$  & $2.60511$  & -       & $66.2497$   & $6.43653$   & -          \\ \hline
\end{tabular}\caption{CIM numerical evaluation for even and odd solutions of $|U(z)|^2$, where we have used $a=10^{-6}$. \label{table}}
\end{table}

\subsection{Correction to Newton's potential in the DSG model}\label{sec:newton}

One of the well-known approaches to studying the corrections in Newton's law is through a Yukawa type interaction between two point masses $M_1$ and $M_2$ separated by a distance $\x$ in four-dimensions \cite{Csaki,Murata,Kapner}.
\begin{gather}
\text{V}(\x)=-G_N\frac{M_1 M_2}{\x}\left[1+ \Delta(\x)\right], \quad \Delta(\x)= \alpha \e^{-\frac{\x}{\lambda}} \label{eq-delta1}
\end{gather}
$\alpha$ and $\lambda$ are adjustable parameters. Ref. \cite{Murata} present a review of these corrections in some theoretical models, while \cite{Kapner} tests the experimental validity of these corrections.

In the braneworlds context, each massive model contributes to the corrections to Newton's potential \cite{RS,Csaki}. The term $\lambda=1/m$, and the amplitude term reads $\alpha(m)=\e^{-\frac{3}{2}A(z)}\hat{U}(z)$ \cite{RS,Csaki},  where $\hat{U}(z)$ is the normalized eigenfunction from Eq \eqref{eq:schro}. $\alpha(m)$ should be applied to the extra-dimensional position where the brane is placed (at $z=0$ \cite{Csaki}). Therefore, the corrections decreases exponentially with both increasing  mass and position. Therefore, for a continuous mass interval, we can express Newton's corrections resulting from the massive modes in a five-dimensional model as an integral over the massive modes
\begin{gather}
  \Delta(\x) \propto\int_{m_1}^{\infty}{\left[\e^{-\frac{3}{2}A(z)}\hat{U}(z)\right]_{z=0}^2\e^{-m\x}} dm \label{eq-delta2}
\end{gather}
 where $\hat{U}(z)$ is the normalized eigenfunction from Eq \eqref{eq:schro} and $m_1$ is the mass of the first excited state.

\begin{figure*}
 \centering
\includegraphics[width=0.75\textwidth]{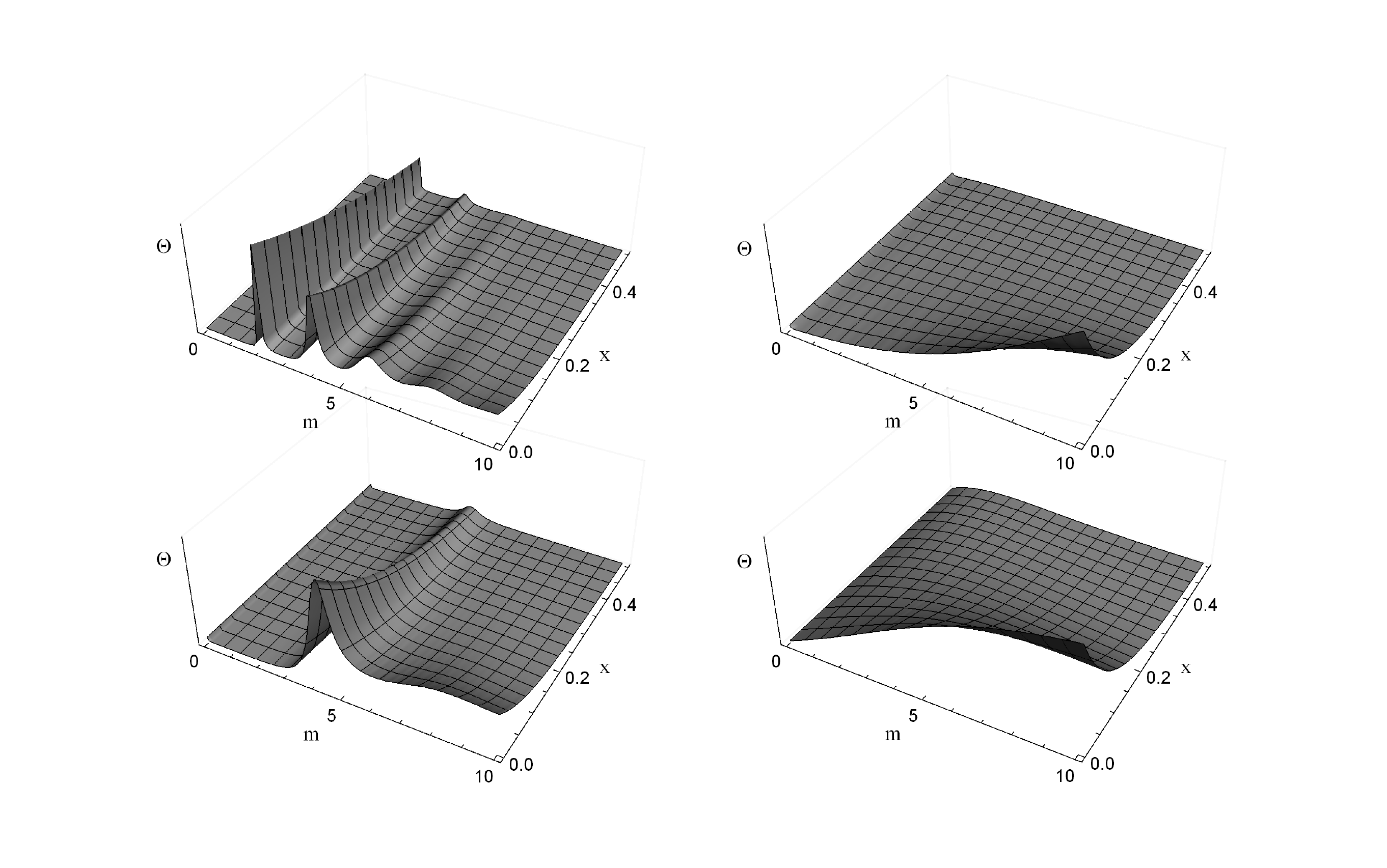}
 \caption{The $\Theta(m,\x,a)$ integrand of $\Delta(\x)$ for $a=10^{-6}$ (top-left panel), $a=10^{-2}$ (bottom-left panel), $a=3.5$ (top-right), and for the SG model (bottom-right).}
 \label{fig:alpha}
\end{figure*}

For the DSG model, we  numerically solve the Eq. \eqref{eq:schro}, considering only the even solutions, as the odd-solutions are null at $z=0$ and do not contribute to the corrections.  We use the mass interval $0.1 \leq m \leq 10$ and the DSG parameters $a=10^{-6}$, $a=10^{-2}$, and $a=3.5$, in addition to the solutions of the SG model. From these massive  eigenfunctions, the plot in Figure  \ref{fig:alpha} represents the integrand $\Theta(m, \x,a)$ of $\Delta(\x)$ from Eq. \eqref{eq-delta2}. When $\x=0$, the profile of $\Theta(m, \x,a)$ shares some similarity with the even solutions resonances for $a=10^{-6}$ and $a=10^{-2}$ in Figure  \ref{fig:ress1a}, decreasing as $\x$ increases. Note that the integrand of $ a = 3.5 $ (DSG model with a single kink) and the integrand of the SG model are similar. Moreover, for smaller values of $a$, we note that the multi-resonance states result in more intense corrections at the origin. The corrections are suppressed with increasing parameter, converging the corrections of the DSG model into SG.

\begin{figure}
 \centering
\includegraphics[width=0.4\textwidth]{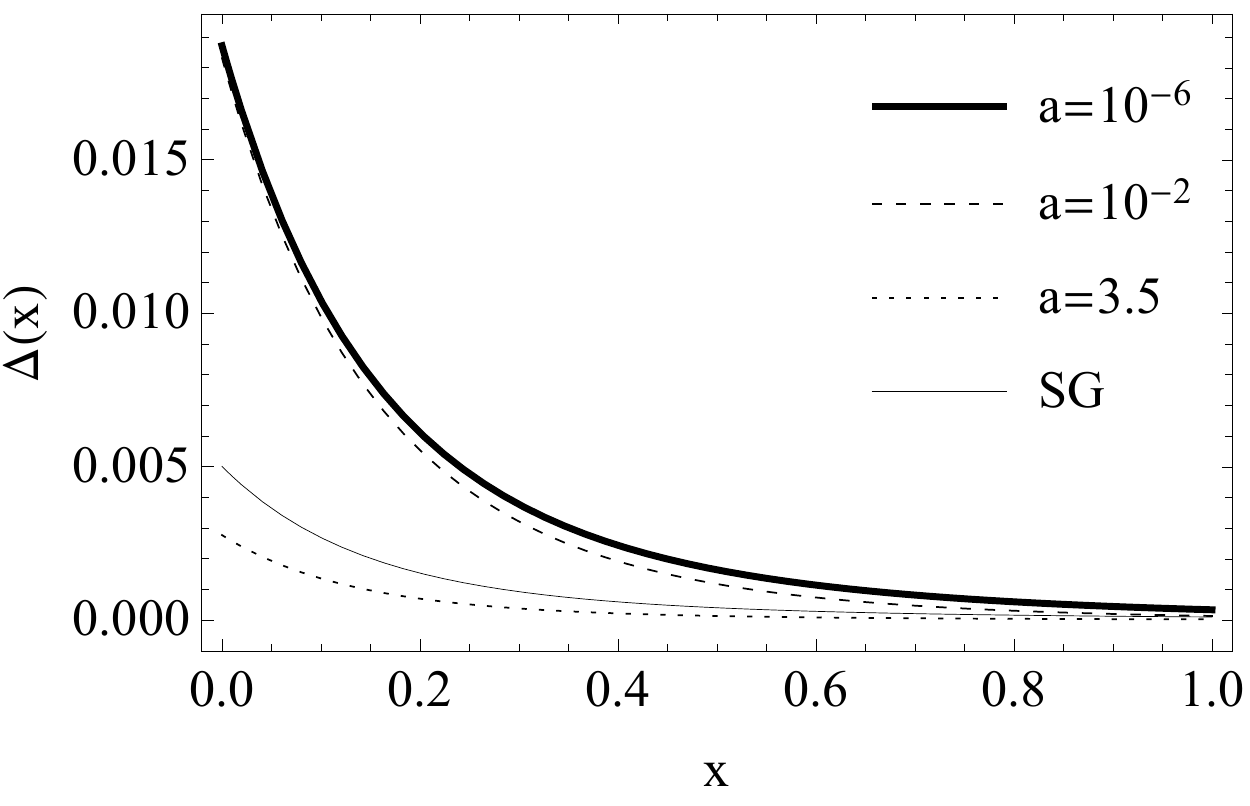}
 \caption{$\Delta(\x)$ corrections to Newton's potential for some  DSG model parameters and the SG model.}
 \label{fig:delta}
\end{figure}

The plot of the $\Delta(\x)$ corrections to Newton's potential are presented in Figure  \ref{fig:delta}. The increase in the parameter $a$ decreases the shape of the corrections. The profiles for $a=10^{-6}$ and $a=10^{-2}$ have similar results, the same occurs for $a=3.5$ and the SG corrections.

\section{Summary}\label{sec.conc}

In this work, we have analyzed a five-dimensional thick brane model generated by a double sine-Gordon potential. We focus our research by considering the evaluation of the  {differential configurational entropy (DCE)} and the corrections to Newton's Law to reveal new aspects concerning the braneworld structure and the gravity localization process.   

Defining the scalar field potential and obtaining the solutions to the equations of motion, we find a degenerate energy density spatial profile.  Furthermore, the system admits a phase transition connecting two classes of domain wall profiles. The transition between the kink solutions to the double bounce structure can be controlled by a parameter included in the scalar field potential. This parameter is also present in the degenerate energy solutions and was used to compute the variation of the DCE.  The  {DCE} evaluation can successfully predict the presence of a critical behavior in the DSG braneworld.  Our results show that there is an energy density profile with minima  {DCE} connected to the raising of an intermediate film between the two interfaces of the kink solution.  {We also found a local minima in the DCE profile related to oscillations on the double sine-Gordon potential.}

Analyzing the massive spectra of the graviton we found a multi-resonance scenario for both even and odd solutions of the massive modes coming from the Schr\"{o}dinger equation.  The resonance modes and their respective lifetimes are also sensitive to the phase transition of the model being confined to the phase corresponding to the two-kink sector.  We also observe that the  {DCE} evaluation can reveal us the phase sector where the resonant modes can exist. The resonance lifetimes are increased for the phase corresponding to the region in the scalar field solution where the two interfaces of the domain wall are separated by a transient state. This sector corresponds to the region at the beginning of the first stage interval of the two-kink solutions. At the phase transition ( {DCE} minimum), there are no resonant modes. 

{The evaluation of the CIMs in terms of the graviton probability density function reveal new aspects concerning the resonances and their respective lifetimes. Both complexity and entropy configurational differential measures present maxima peaks at masses corresponding to the resonances.  Conversely, for solutions of the corresponding plane waves $m^2>\mathcal{V}_{max}$, the CIMs acquire the lowest values found. These results indicate these informational measures as new methods to approach the graviton localization in braneworlds, as they can be used to predict the existence of resonant modes.}

Finally, we performed corrections to Newton's potential in the DSG varying by the parameter $a$. We note that  Newton's potential shares some similarities with the resonant profiles. We conclude that smaller $a$ exhibits longer lifetimes and multi-resonance states, which have more significant contributions to the corrections. As the parameter $a$ increases, the resonances vanish, and the single kink profile is achieved, taking the DSG model into a form similar to that of the SG model.  The most significant corrections to the Newtonian potential reside in the phase characterized by the two-kink solutions. 

\section{Acknowledgments}
This work was supported by the Brazilian
agencies Coordena\c{c}\~ao de Aperfei\c{c}oamento de Pessoal de
N\'{i}vel Superior (CAPES), the Conselho Nacional de Desenvolvimento
Cient\'{i}fico e Tecnol\'ogico (CNPq), and Fundacao Cearense de apoio ao Desenvolvimento Cientifico e Tecnologico (FUNCAP).


\end{document}